\documentclass{sig-alternate}

\usepackage{psfrag}
\usepackage{algorithm}

\usepackage[noend]{algpseudocode}
\usepackage{rotating}

\newcommand{\<}{\langle}
\renewcommand{\>}{\rangle}

\newcommand{\reals}{{\mathds R}}

\def\cC{{\cal C}}
\def\cR{{\cal R}}

\def\l|{\left|\left|}
\def\r|{\right|\right|}
\def\R{\mathbb R}
\def\P{\mathbb P}

\def\prob{{\mathbb P}}

\def\cL{{\cal L}}

\renewcommand{\theenumi}{{\roman{enumi}}}

\def\prob{{\mathbb P}}

\def\reals{{\mathbb R}}

\def\<{\langle}
\def\>{\rangle}

\def\P{{\sf P}}
\def\AUC{{\sf AUC}}

\def\normal{{\sf N}}
\def\gossip_pca{{\sc Gossip PCA}}

\def\hM{\hat{M}}
\def\one{1}

\def\cO{{\mathcal{O}}}

\def\Train{{\sf Train}}
\def\Test{{\sf Test}}
\newcommand{\T}[1]{{\sf T#1}}
\newcommand{\TP}[1]{{\sf TP#1}}
\newcommand{\TPR}[1]{{\sf TPR#1}}
\def\alg{{\rm Alg}}

\begin{document}
\conferenceinfo{CAMRa2011,} {October 27, 2011, Chicago, IL, USA.}
\CopyrightYear{2011}
\crdata{978-1-4503-0825-0}
\title{Identifying Users From Their Rating Patterns}

\numberofauthors{1}

\author{
\alignauthor Jos\'e Bento$^*$, Nadia Fawaz$^\dagger$, Andrea Montanari$^{* \ddagger}$, and Stratis Ioannidis$^\dagger$\\
\affaddr{$^{*}$Department of Electrical Engineering, Stanford University,\\
$^\ddagger$Department of Statistics, Stanford University,\\
 $^\dagger$Technicolor
}\\
\email{jbento@stanford.edu, nadia.fawaz@technicolor.com,\\
 montanari@stanford.edu, stratis.ioannidis@technicolor.com}
}

%\numberofauthors{4}
%\author{
%% 1st. author
%\alignauthor
%	Jose Bento\\
%        \affaddr{Department of Electrical Engineering}\\
%        \affaddr{Stanford University}\\
%        \email{jbento@stanford.edu}
%% 2nd. author
%\alignauthor
%	Nadia Fawaz\\
%       \affaddr{Technicolor}\\
%       \email{nadia.fawaz@technicolor.com}
%\and
%% 3rd. author
%\alignauthor
%	Stratis Ioannidis \\
%       \affaddr{Technicolor}\\
%       \email{stratis.ioannidis@technicolor.com}
%% 4th author
%\alignauthor
%	Andrea Montanari\\
%       \affaddr{Department of Electrical Engineering}\\
%       \affaddr{Department of Statistics}\\
%       \affaddr{Stanford University}\\
%       \email{montanari@stanford.edu}
%}

% There's nothing stopping you putting the seventh, eighth, etc.
% author on the opening page (as the 'third row') but we ask,
% for aesthetic reasons that you place these 'additional authors'
% in the \additional authors block, viz.
% Just remember to make sure that the TOTAL number of authors
% is the number that will appear on the first page PLUS the
% number that will appear in the \additionalauthors section.

\maketitle

\begin{abstract}
This paper reports on our analysis of the 2011 CAMRa Challenge dataset
(Track 2) for context-aware movie recommendation systems. The train dataset
comprises $4 \: 536 \: 891$ ratings provided by $171 \: 670$ users on $23 \: 974$ movies, as well as the
household groupings of a subset of the users. The test dataset
comprises $5 \: 450$ ratings for which the user label is missing, but the
household label is provided. The  challenge required to identify the
user labels for the ratings in the test set.

Our main finding is that temporal information  (time labels of the
ratings) is significantly more useful for achieving this objective
than the user preferences (the actual ratings). Using a model that
leverages on this fact, we are able to identify users within a known
household with an accuracy of approximately $96\, \%$
(i.e. misclassification rate around $4\,\%$).
\end{abstract}

\category{G.3.}{Probability and Statistics}{Correlation and regression analysis}


%
%=========================================================================
%
\section{Introduction}

The incorporation of contextual information
is likely to play an ever-increasing role  in recommendation systems
because of the broad availability of such information, and the need for more accurate
systems. Among sources of contextual information, the social structure of a given
pool of users is particularly interesting in view of the potential
convergence between online social networks and recommendation systems.

In this paper we investigate the relation between social structure and
users behavior within a recommendation system, through the analysis
of the CAMRa 2011 dataset (Track 2). Our results are summarized in Table \ref{table:bestresults}.

In the remainder of this section we describe the challenge data set, we explain the performance metrics used, we give an overview of the algorithms we propose and their corresponding results, and finally we give a short overview of related work.

\begin{table}
\begin{center}
\begin{small}
\begin{tabular}{r|c|c|c|c|}
\cline{2-5}
 & Any size & Size 2  & Size 3  & Size 4   \\
\hline
\multicolumn{1}{|r|}{Misclassification rate}& 0.0406 & 0.0413 & 0.0268 & 0.0463\\
\hline
\end{tabular}
\end{small}
\caption{Best misclassification rates obtained for the challenge data set (Track 2). We report the average misclassification rate over all households, average over all households of size 2, of size 3 and of size 4 respectively.}\label{table:bestresults}
\end{center}
\end{table}

\subsection{Description of the data set}

The training data consists of a collection of $4 \; 536 \; 891$ ratings. Each entry
(rating) takes the form
\begin{eqnarray}
(i,j,M_{ij},t_{ij}).
\end{eqnarray}
Here $i\in [m]$ (with $m=171 \: 670$) is a user ID, $j\in [n]$ (with
$n=23 \: 974$) is a movie ID, $M_{ij}$ (with $0\le M_{ij}\le 100$) is the
rating provided by user  $i$ on movie $j$, and $t_{ij}$ is the
time-stamp of that rating. (Throughout this paper we denote by
$[N] =\{1,\dots,N\}$ the set of first $N$ integers.)
We denote by $E\subseteq [m]\times [n]$ the subset of user-movie
pairs for which a rating is available.

The training data also includes information about the household structure
of a subset of users. This provided in the form of $290$ household-composition tuples
\begin{eqnarray}
(H,i_1,\dots,i_k)\, .
\end{eqnarray}
Here $H$ is a household ID, and $i_1,\dots,i_L$ are the IDs of users
belonging to household $H$. The number $L$ of users in the same
household varies between $2$ and $4$. We will write $i\in H$ to
indicate that user $i$ belongs to household $H$. For instance, given
the above tuple, we know that $i_1,\dots,i_L\in H$.

The test data comprises $5 \:450$  tuples of the form
\begin{eqnarray}
 (H,j,M_{Hj}, t_{Hj})\, ,
\end{eqnarray}
whereby $H$ is an household ID, $j$ is a movie ID, $M_{Hj}$ is a
rating provided by one of the users in $H$ for movie $j$, and
$t_{Hj}$ is the corresponding time-stamp. The challenge Track 2 requires to
infer the user $i\in H$ that actually provided these ratings.

In the following, we denote by $\Train$ the train set, and by
$\Test$ the test set. % (i.e. with a slight abuse of notation, either the list
%of tuples introduced above, or of indices for those tuples).
%
\subsection{Performance metrics}

Of the $290$ households, the vast majority, namely $272$, is formed by $2$ users,
while $14$ include $3$ users, and only $4$ are formed by $4$ users. As a consequence of this,
a purely random inference algorithm achieves an average misclassification rate over all households that
is slightly above $50\, \%$ (indeed, approximately $0.511$). The same random inference algorithm achieves an average misclassification rate of $50\, \%$ over households of size 2, of $66\, \%$ over households of size 3 and $75\, \%$ over households of size 4. This performance
provides a baseline for the algorithms developed in this paper.

As a performance metric we will use standard ROC variables (true
positive rate and one minus false positive rate). More precisely,
given a household with two users $i=1$ and $i=2$, we let $\T{1}$ and $\T{2}$ be
the total number of entries in \Test, that correspond to user $1$ and user
$2$ respectively while, $\TP{1}(\alg)$, $\TP{2}(\alg)$ are the the number of those
entries assigned by algorithm $\alg$ to $1$ and $2$. Then the
corresponding true positive rates are
\begin{eqnarray}
\TPR{1}(\alg) = \frac{\TP{1}(\alg)}{\T{1}} \, ,\;\;\;\;\;\;\;\TPR{2}(\alg) =
\frac{\TP{2}(\alg)}{\T{2}} \, .
\end{eqnarray}
Notice that $\TPR{2}(\alg)$ is equal to one minus the false positive
rate in predicting $1$, so these are the usual ROC variables.
This definition is generalized in the obvious way in the case of $3$-
and $4$-user households. 

The total misclassification rate per household $H$ is defined as follows in terms of the above
quantities (always considering $2$-user households but easily generalized)
\begin{eqnarray}
\P(\alg,H)\equiv 1-\frac{\TP{1}(\alg)+\TP{2}(\alg)}{\T{1}+\T{2}}\, .
\end{eqnarray}
We define $\P$ to be the average of $\P(\alg,H)$ over all households. We also compute the average of $\P(\alg,H)$ over households of size 2 only, of size 3 only and size 4 only. We denote these values by $\P_{2}$, $\P_{3}$ and $\P_{4}$ respectively.

In order to obtain a $2$-dimensional ROC curve, we
will plot the true positive rate for -say- user $1$ against the true
positive rate for the union of users $2$ and $3$.

%
%************************************************************************
%
\subsection{Overview of algorithms and results}

We will consider three classes of methods that incorporate increasing
amounts of contextual information:

\noindent {\bf 1.} \emph{Low-rank approximation},  cf. Section
\ref{sec:LowRank}, provides an effective tool to embed the collection
of movies and users at hand, within a low-dimensional latent space
$\reals^r$, $r\ll m,n$. A high rating provided by user $i$ on movie
$j$ corresponds to latent space vectors with large inner product.
We use the latent vectors associated with users within the same
household to infer which user rated a certain movie, by selecting the
latent vector whose inner product with the movie vector best
reproduces the observed rating.
Generalizing \cite{Kor08}, we extend these models to include temporal
variability, in both users' and movies' latent vectors. If our temporal
units are the 12 months of the year, the resulting
model achieves an overall misclassification rate $\P \approx 0.3735$.

\noindent {\bf 2.} The second group of methods, cf. Section
\ref{sec:Time},  makes a crucial use of \emph{temporal patterns} in
the users rating behavior. Indeed, our single most striking discovery
is that different users within the same household exhibit very well
separated viewing habits. These habits are clearly demonstrated by comparing
the distribution of ratings across the days of the week for two users
in the same household. For a large number of households, these
distributions have almost disjoint support.
A simple algorithm that uniquely uses the day of the
week to infer the user identity, achieves a misclassification
rate $\P\approx 0.1154$.  We also discuss a generative model
which incorporates both ratings (through low-rank approximation) and
temporal patterns, achieving $\P\approx 0.0950$.

\noindent {\bf 3.} Section \ref{sec:Unified} proposes a \emph{unified
  framework} based on binary classification to exploit latent
space information as well as temporal information, and additional
contextual information. The binary classification `module' we use is
regularized logistic regression, but could be replaced by a number of
equivalent methods. By using composite feature vectors including
several types of information, we achieve $\P\approx 0.0406$.
%
%************************************************************************
%
\subsection{Related work}

Several aspects of our investigation confirm claims of earlier
work, such as the usefulness of low-rank approximation \cite{BK07,KBV09} and the importance of accounting for temporal evolution \cite{KBV09,Gantner}.
%the challenge posed by accounts shared by multiple users \cite{}.
At the same time, the present dataset allows us to provide
striking evidence of these two points. Furthermore, the precise
form of temporal patterns and their extraction in the form of weekly
and daily habits is novel and extremely powerful.

The importance of the time of day as context for recommendations has been noted in the past, \emph{e.g.}, in  recommending music tracks \cite{music2,music1}. Our most striking finding is that, in the challenge dataset, users within a given household tend to view and rate movies at different times of the day and different days of the week. Thus, time is an important factor not only in recommendations but also in user identification.

%
%{\bf [A: Please add whatever references you think useful]}

%=========================================================================
%
\section{Low-rank approximation}
\label{sec:LowRank}

This section consists of three parts, dealing respectively with rating prediction from a training set, rating classification in a test set, and evaluation of the misclassification rate on the challenge data set.
We first propose two collaborative filtering methods, based on low-rank matrix completion, to predict the missing ratings in a training set. The first method relies only on the ratings provided in the training set to predict the missing ratings. The second method also factors in the context by taking into account the temporal information in the training set. We then turn our attention to the test set, containing household ratings, and use the aforementioned prediction models to identify which user in a household provided a given rating in the test set. Finally, we evaluate our methods on the challenge dataset, and provide empirical results in terms of misclassification rate and ROC curve.

Throughout this section, we denote by $x \sim U[a,b]$ a random variable $x$ uniformly distributed in $[a,b]$.
For $x,y \in \R^n$, $\<x,y\>= x^T y =\sum_{\ell=1}^{n} x_\ell
y_{\ell}$ denotes the usual inner product, and $\|x\|^2 = \<x,x\>$.
 For  $M \in \R^{m\times n}$,  $\| M \|_F$ is its Froebenius norm.
We let $\one_{n}=[1,\ldots,1]^T$, and $I_n$ be the identity matrix of
size $n$.
%Given a matrix $A\in \R^{r \times n}$, two column vectors $x,y \in
%\R^n$, and two real numbers $\alpha, \beta \in \R$,
%we define the functions $g(A,x,\alpha)=(A A^T + \alpha I_r)^{-1} A
%x$, and $h(A,x,y,\alpha,\beta)=(A A^T +
%\alpha I_r)^{-1} (Ax + \beta y)$.

%--------------------------------------------------%
\subsection{Simple low-rank approximation}\label{sec:LR}
%--------------------------------------------------%

\subsubsection{Model}

A simple low rank model is obtained by approximating the matrix of
ratings $M\in\reals^{m\times n}$ by a low-rank matrix $\hM= UV^T + Z\one_n^T$,
where matrix $U = [u_1|\cdots|u_m]^T$ is of size $m \times r$, matrix $V=[v_1|\cdots|v_n]^T$ is of size $n\times r$, and the column vector $Z=[z_1,\ldots,z_m]^T$ is of length $m$. Each vector $u_i\in\reals^r$ is associated with a user $i\in [m]$, and each vector $v_j\in\reals^r$ corresponds to a movie $j\in [n]$. The column vector $Z$ models the rating bias of each user.
Matrices $U$, $V$ and $Z$ are found by minimizing the following
regularized empirical $\ell_2$ loss
\begin{equation}\label{eq:LRCost}
\begin{split}
\cC(U,V,Z) \equiv
&\frac{1}{2}\!\sum_{(i,j)\in E}\!\!\big(M_{ij}\!-\!\<u_i,v_j\>\! -\!z_i\big)^2 \\
& \!+\! \frac{\lambda}{2} \|U \|_F^2
  \!+\! \frac{\lambda}{2} \| V \|_F^2.
\end{split}
\end{equation}

\subsubsection{Alternate minimization}\label{sec:LR-Alg}

%\algsetup{linenosize=\tiny}
\begin{algorithm}[tpb]
\caption{Low rank approximation}\label{alg:LR}
{ \fontsize{7}{7}\selectfont
\begin{algorithmic}[0]
\Procedure{Initialization}{}%\Comment{ }
\State $\forall  (i,j) \in [m] \times [r], \quad u_{ij}^{(0)} \sim \frac{U[0,1]}{\sqrt{m}}$
\State $\forall  (i,j) \in [r] \times [n], \quad v_{ij}^{(0)} \sim \frac{U[0,1]}{\sqrt{n}}$
\State $\forall  i \in [m], \quad z_{i}^{(0)} = 50 $
\EndProcedure
\Statex
\Procedure{Iterations}{$K$}
\For{$k =  1 \ldots K$}
    \For{$i =  1 \ldots m$}
        \State $ u_i^{(k)} = g\left(V_{E_i}^{(k-1)}, \: M_{i\,E_i}^T-1_{|E_i|}z_i^{(k-1)}, \: \lambda \right)$
    \EndFor
    \For{$j =  1 \ldots n$}
        \State $v_j^{(k)} = g\left({U_{F_j}^{(k)}}^T, \: M_{F_j\,j} - z_{F_j}^{(k-1)}, \: \lambda \right)$
    \EndFor
    \For{$i =  1 \ldots m$}
        \State $z_i^{(k)}= g\left(1_{|E_i|}^T, \: M_{i\,E_i}^T - {V_{E_i}^{(k)}}^T u_i^{(k)} , \: 0 \right) $
    \EndFor
\EndFor
\EndProcedure
\Statex
\State \textbf{Return} $(U^{(K)},V^{(K)},Z^{(K)})$
\end{algorithmic}
}
\end{algorithm}

The cost function (\ref{eq:LRCost}) is non convex, but several
iterative minimization methods have been developed with excellent
performances in practical settings
\cite{SRJ05,RS05,HH09,MJD09powerlaw,WYZ10}.
Performances guarantees for algorithms of this family were proved in
\cite{KMO09,KMO09noise}, under suitable assumptions on the matrix
$M$. Alternative approaches based on convex relaxations have been
studied in \cite{CaR08,Gross09}.

In this paper we adopt a simple alternate minimization  algorithm
(see e.g. \cite{Kor08,HH09} for very similar algorithms).
Each iteration of the algorithm consists of three steps: in the first step, $V$ and $Z$ are fixed, and $U$ is updated by minimizing (\ref{eq:LRCost}); then $U$ and $Z$ are fixed, and $V$ is updated; finally, $U$ and $V$ are fixed and $Z$  updated.
A pseudocode for the algorithm is presented in Algorithm~\ref{alg:LR}.
The algorithm stops after $K$ iterations, and returns the triplet
$(U,V,Z)$.

Since the cost  (\ref{eq:LRCost}) is separately quadratic in each of
$U$, $V$ and $Z$, each of the steps can be performed by matrix
inversion.  In fact, the problem presents a convenient separable
structure. For instance, the problem of minimizing over $U$ is
separable in $u_1$, $u_2$, \dots, $u_m$.
Minimizing $\cC(U,V,Z)$ over a vector $u_i$ is equivalent to a Ridge regression in $u_i$, whose exact solution is given by
\begin{equation}
u_i = (V_{E_i} {V_{E_i}}^T + \lambda I_r)^{-1} V_{E_i} ( M_{i\,E_i} - z_i 1_{|E_i|}^T)^T \mbox{, }
\end{equation}
where $E_i=\{j\in[n] | (i,j) \in E \}$, $M_{i\,E_i}=[m_{ij}]_{j\in E_i}\in\R^{1\times |E_i|}$, and $V_{E_i}=[v_j]_{j\in E_i}\in\R^{r \times |E_i|}$.
In order to concisely represent this basic update, we define the function
$g$ as follows.
Given a matrix $A\in \R^{r \times n}$, a column vector $x \in
\R^n$, and a real number $\alpha, \beta \in \R$, we  let
$g(A,x,\alpha)\equiv(A A^T + \alpha I_r)^{-1} A x$. The above update
then reads $u_i= g(V_{E_i} , \: M_{i\,E_i}^T-1_{|E_i|}z_i , \: \lambda)$. Define $F_j=\{i\in[n] | (i,j) \in E \}$. 
proceeding analogously for the minimization over $V$ and $Z$, we obtain  Algorithm~\ref{alg:LR}.

%--------------------------------------------------%
\subsection{Low rank approximation with\\ time-dependent factors}
%--------------------------------------------------%
\label{subsec:LowRankTD}

In this section, we extend the previous low-rank prediction model to account for temporal information.

\subsubsection{Model}

In this model, we bin time into $T$ bins of equal duration, indexed by $ b \in \{1,\dots,T\}$.
Given that user $i$ rates movie $j$ at time $t_{ij}$, we denote by $b(t_{ij})\in [T]$
the unique bin index for the observed rating of the pair $(i,j)$.
%Given a bin index $t\in [T]$, we denote by $E(t)=\{(i,j)\in E | b(t_{ij}) = t\}$ the subset of user-movie pairs $(i,j)$ for which the rating was provided at a time $t_{ij}$ in bin $t$.

Let $M \in \R^{m \times n \times T}$ be the three-dimensional rating
tensor whose entry $M_{ij}(b)$ represents the rating that
user $i\in[m]$ would give to movie $j\in[n]$ at a time in bin
$b\in[T]$. The matrix $M(b)\in \R^{m\times n}$ represents the rating matrix in
bin $b$.  From a training set of observed ratings $\{M_{ij}(b) | (i,j)
\in E \}$, we predict the missing ratings by approximating
each matrix $M(b)$, $b\in[T]$ by a low rank matrix $\hM(b) = U(b)
V(b)^T + Z(b) 1^T_n$.
This is a natural extension of the model in
Section~\ref{sec:LR}. Matrices $U(b)\in\R^{m \times r}$, $V(b) \in \R
^{n \times r}$ and $Z(b) \in \R^{m \times 1}$ are stacked in the tensors
$U \in \R^{m \times r \times T}$, $V \in \R^{r \times n \times T}$ and $Z \in \R^{m \times 1 \times T}$ respectively.
We obtain the tensors $(U,V,Z)$ by minimizing the following
regularized $\ell_2$ loss
\begin{equation}\label{eq:TD-LRCost}
\begin{split}
& \cC(U,V,Z)  \equiv\cR_{\lambda,\xi_u}(U) +\cR_{\lambda,\xi_v}(V) +\cR_{0,\xi_z}(Z) +\\
&  \frac{1}{2}\!\!\sum_{(i,j)\in E}\!\!\!\! \left( M_{ij}(b(t_{ij}))\! -\!\<u_i(b(t_{ij})),v_j(b(t_{ij}))\>\!-\!z_i(b(t_{ij})) \right)^2\!\!\!,  %\nonumber
\end{split}
\end{equation}
where the regularization terms are of the form
\begin{equation}\label{eq:TD-LRreg}
\begin{split}
\cR_{\lambda,\xi}(U)
&=  \frac{\lambda}{2} \sum_{b=1}^T\! \|U(b)\|_F^2
+\frac{\xi}{2} \sum_{b=1}^{T-1}\! \|U(b+1)\!-\!U(b)\|_F^2\, .
\end{split}
\end{equation}
Each regularization function consists of two terms: the first term is
an $\ell_2$  regularization for shrinkage, while the second term
promotes smooth time-variation.
Note that by setting the number of bins to $T=1$,
this model reduces to the time-independent model described in
Section~\ref{sec:LR}.
The same happens by letting $\xi_u,\xi_v,\xi_z\to\infty$.

\subsubsection{Alternate minimization}

\begin{algorithm*}[t!]
\caption{Time-dependent low rank approximation}\label{alg:TD-LR}
{\fontsize{7}{7}\selectfont
\begin{algorithmic}[0]
\Procedure{Initialization}{}%\Comment{ }
\State $\forall  (i,j,b) \in [m] \times [r] \times [T], \quad {u_{ij}(b)}^{(0)} \sim \frac{U[0,1]}{\sqrt{m}}$
\State $\forall (i,j,b) \in [r] \times [n] \times [T], \quad {v_{ij}(b)}^{(0)} \sim \frac{U[0,1]}{\sqrt{n}}$
\State $\forall  (i,b) \in [m] \times [T], \quad {z_{i}(b(t))}^{(0)}  = 50 $
\EndProcedure
\Statex
\Procedure{Iterations}{$K,T$}
\For{$k =  1 \ldots K$}
    \For{$b =  1 \ldots T$}
        \For{$i =  1 \ldots m$}
            \State  $ u_i(b)^{(k)} = h \left(V_{E_i(b)}^{(k-1)},\: M_{i\,E_i(b)}^T -  1_{|E_i(b)|}z_i(b)^{(k-1)}, \: {u_i(b+1)}^{(k-1)}+{u_i(b-1)}^{(k)},\: \lambda+2\xi_u, \: \xi_u\right)$
        \EndFor
        \For{$j =  1 \ldots n$}
            \State $v_j(b)^{(k)} = h\left({U_{F_j(b)}^{(k)}}^T,\: M_{F_j(b)\,j} - z_{F_j}(b)^{(k)}, \: {v_j(b+1)}^{(k-1)}+{v_j(b-1)}^{(k)} ,\: \lambda+2\xi_v, \: \xi_v\right)$
        \EndFor
        \For{$i =  1 \ldots m$}
            \State $z_i(b)^{(k)}=h\left(1_{|E_i(b)|}^T,\: {M_{i\,E_i(b)}}^T - {V_{E_i(b)}^{(k)}}^T {u_i(b)}^{(k)}, \: z_i(b+1)^{(k-1)}+z_i(b-1)^{(k)} ,\: 2\xi_z, \: \xi_z\right)$
        \EndFor
    \EndFor
\EndFor
\EndProcedure
\Statex
\State \textbf{Return} $(U^{(K)},V^{(K)},Z^{(K)})$
\end{algorithmic}
}
\end{algorithm*}

In order to minimize the cost function (\ref{eq:TD-LRCost}), we
generalize the alternate minimization algorithm of Section
\ref{sec:LR-Alg}. Namely we cycle over the time bin index $b$ and, for
each $b$, we sequentially  minimize over
$U(b)$, $V(b)$ and $Z(b)$, while keeping $U(b')$, $V(b')$ and $Z(b')$,
$b'\neq b$ fixed.
As before, each of these three minimization problems is quadratic
and hence solvable efficiently.
Further, each of these quadratic problems is separable across user
indices (for minimization over $U$ and $Z$) or movie indices (for
minimization over $V$). On the other hand, it is not separable across
time bins because of the second term in the regularization function, cf.
Eq.~(\ref{eq:TD-LRreg}). As a consequence, the update steps change
somewhat.
Consider --to be definite-- the minimization over $U$. A
straightforward calculation yields the following expression for the
minimum over $u_i(b)$, when all other variables are kept constant
\sloppy
\begin{equation*}
\begin{split}
& u_i (b) =  \left(V_{E_i(b)} {V_{E_i(b)}}^T + (\lambda + 2 \xi_u) I_r \right)^{-1} \times \\
& \!\!\! \left( V_{E_i(b)} ( M_{i\,E_i(b)} - z_i(b) 1_{|E_i(b)|}^T)^T + \xi_u
  \left(u_i(b+1)+u_i(b-1)\right)  \right)
\end{split}
\end{equation*}
where we assumed $b\in \{2,\dots,T-1\}$ (the boundary cases $b=1,T$
yield slightly different expressions).
Defining $h(A,x,y,\alpha,\beta)=(A A^T + \alpha I_r)^{-1} (Ax + \beta
y)$,  the above can be written as
$u_i(b)= h(V_{E_i(b)},\: M_{i\,E_i(b)}^T -  1_{|E_i(b)|}z_i(b), \:
  u_i(b+1)+u_i(b-1) ,\: \lambda+2\xi_u, \: \xi_u)$.

\fussy
Analogous expressions hold  for minimization over $z_i(b)$ and
$v_j(b)$. A complete pseudocode is provided in Algorithm~\ref{alg:TD-LR}.

\subsection{Household rating classification and results}

For each entry in the test set, the goal is to identify which user in
the household provided the rating.
In this section, our approach uses the rating and the corresponding time-stamp provided within the test
set, and the low rank model obtained from the training set. Given
a rating $M_{Hj}$ within household $H =\{i_1,\dots i_L\}$, the simplest idea is to attribute the
rating to the user $i\in H$ for which the predicted rating is closest
to $M_{Hj}$. In other words, we return $\arg\min_{i\in H}|M_{Hj}-\hM_{ij}(b(t_{Hj}))|$.

In order to explore the tradeoff between precision and accuracy through an ROC curve, we slightly generalize  this rule by introducing a parameter $\alpha\ge
0$, and proceed as follows.
\begin{enumerate}
\item 
First, for each user $i \in H$, we compute the difference:  $|M_{Hj}-\hM_{ij}(b(t_{Hj}))|$.
\item Consider the first user $i_1 \in H$. If $$\alpha
  |M_{Hj}-\hat{M}_{i_1 j}(b(t_{Hj}))| < \min_{i\in H \backslash
    i_1} |M_{Hj}-\hat{M}_{ij}(b(t_{Hj}))|,$$ we conclude that user
  $i_1$ provided the household rating $M_{Hj}$.
Otherwise, we conclude it was some other user in the household.
%, and repeat the comparison for each user in the household.
\end{enumerate}

\subsubsection{Parameter selection and results}
\label{sec:LowRankParam}

We will limit ourselves to discussing the results obtained with
time-dependent factorization, since this method leads to more accurate
predictions, and it subsumes the time-independent approach as a
special case.

\begin{figure}[t!]
\begin{center}
\psfrag{Number of iterations K}[cc][cc]{{\footnotesize Number of iterations $K$}}
\psfrag{Misclassification rate}[cc][cc]{{\footnotesize Misclassification rate}}
\psfrag{T=12, r=10, \lambda=1, \xi_u=10, \xi_v=\xi_z=40}[cc][cc]{{\scriptsize $T=12$, $r=10$, $\lambda=1$, $\xi_u=10$, $\xi_v=\xi_z=40$}}
\psfrag{T=1, r=10, \lambda=1}[cc][cc]{{\scriptsize $T=1$, $r=10$, $\lambda=1$}}
\psfrag{T=12, r=5, \lambda=1, \xi_u=10, \xi_v=\xi_z=40}[cc][cc]{{\scriptsize $T=12$, $r=5$, $\lambda=1$, $\xi_u=10$, $\xi_v=\xi_z=40$}}
\psfrag{T=5, r=10, \lambda=1, \xi_u=10, \xi_v=\xi_z=40}[cc][cc]{{\scriptsize $T=5$, $r=10$, $\lambda=1$, $\xi_u=10$, $\xi_v=\xi_z=40$}}
\psfrag{T=5, r=10, \lambda=1, \xi_u=40, \xi_v=\xi_z=40}[cc][cc]{{\scriptsize $T=5$, $r=10$, $\lambda=1$, $\xi_u=40$, $\xi_v=\xi_z=40$}}
\includegraphics[width=0.95\columnwidth]{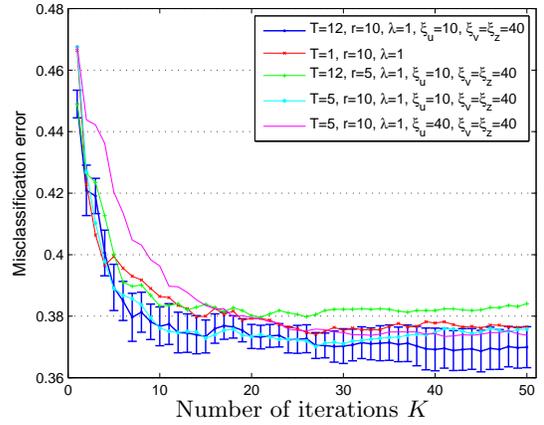}
\caption{Average misclassification rate vs. number of iterations $K$,  for different values of parameters.} % $(T,r,\lambda,\xi_u,\xi_v=\xi_z=40)$.}
\label{fig:LR-PredErr}
\end{center}
\end{figure}

We evaluated the accuracy  through cross-validation for several choices of the regularization parameters.
Figure~\ref{fig:LR-PredErr} shows the average misclassification rate versus the number of iterations for various values of parameters.
The misclassification rate is close to $37\%$, and seems to become
stable after about $50$ iterations. We thus fixed $K=50$, and selected
the following values of parameters by minimizing the misclassification rate:
number of bins $T=12$; rank $r=10$; regularization parameters
$\lambda=1$, $\xi_u=10$, $\xi_v=\xi_z=40$.
Let us emphasize that we did not perform an exhaustive search over all
sets of possible values,
 which could lead to further improvements.

\sloppy
The results in Figure~\ref{fig:LR-PredErr} were obtained by
random-subsampling cross-validation. We averaged
over $5$ different splits of the dataset into training set and test set. In each split, the test set was selected by randomly hiding
approximately $4\%$ of the data of each household. The curves obtained with the original training and
test sets provided in the challenge are close to the
ones in Figure~\ref{fig:LR-PredErr}. Our cross validation procedure is more reliable from a statistical point of
view. We will keep to this procedure for the rest of the paper and
only mention eventual discrepancies with respect to the original split in test
and training set provided in the challenge.

\begin{figure}[t!]
\begin{center}
\psfrag{TPR1}[cc][cc]{{\footnotesize $\TPR1$}}
\psfrag{TPR2}[cc][cc]{{\footnotesize $\TPR2$}}
\includegraphics[width=0.95\columnwidth]{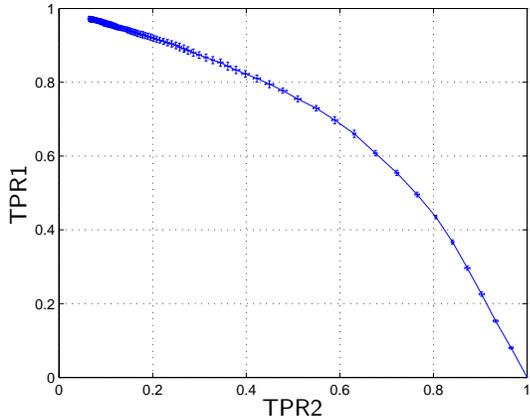}
\caption{TPR of user 1 in each household vs. TPR of any other user.\vspace*{-0.5cm}}
\label{fig:LR-ROC}
\end{center}
\end{figure}

 Figure~\ref{fig:LR-ROC} shows the ROC curve achieved by
the present classification method, for varying $\alpha$.
Each point of the curve corresponds to the average of the pair
$(\TPR1(\alpha),\TPR2(\alpha))$ over all households in a
$(\Train, \Test)$ pair, itself averaged over all $(\Train, \Test)$ pairs
(splits). Bars show the standard deviation from the mean over
different $(\Train,\Test)$ splits.\\
\fussy

%\begin{figure}[t]
%\includegraphics[width=0.98\columnwidth]{prediction_error_using_low_rank_model_diff_val_parameters_with_mean_plus_error_bar.eps}
%\caption{Average prediction error vs. number of iterations for different sets of parameters}
%\label{fig:LR-PredErr}
%\end{figure}

%
%=========================================================================
%
\section{Temporal signatures}
\label{sec:Time}

Although our matrix factorization model captures the evolution of user
and movie profiles throughout the $12$-month
period of the dataset, it does not make direct use of the rating
time-stamp in order to classify ratings within a household. The
time-stamp is only used indirectly, namely to compute the predicted
ratings $\hM_{ij}$.

On the other hand, temporal behavior ---especially weekly behavior--- appears to be extremely useful in
distinguishing users within the same household. Household
members  exhibit distinct temporal patterns in their viewing
habits. Rather than viewing movies together,
in many households users consistently rate movies  at different days of the week.

 \sloppy As a result, the day of the week on which a movie is rated provides a
  surprisingly good predictor of the user who watched it. We exploit
  this finding below, and  propose a generative  model that incorporates the day of the week as well as the movie rating.

\fussy

%\begin{figure}[tbp]
%\includegraphics[width=0.98\columnwidth]{ROC_curve_for_prediction_using_low_rank_lambda_10_lambdaM_40_xi_40_gamma_1_rank_10_50_iterations_mean_plus_error.eps}
%\caption{TPR of user 1 in each household vs. TPR of any other user}
%\label{fig:LR-ROC}
%\end{figure}

\subsection{Temporal patterns in user behavior}\label{sec:temporal}

Clear temporal patterns emerge when considering the day of the week
on which ratings are given.
Most importantly,  the temporal patterns in the viewing behavior of
members of the same household turn out to be very well separated.

\begin{figure}[t]
\psfrag{Number of rating events}[bc][tc]{{\scriptsize Number of rating events}}
\psfrag{Day of the week}[tc][bc]{{\scriptsize Day of the week}}
\psfrag{Frequency of viewings along the week for users of household 1}[bc][tc]{{\scriptsize household 1}}
\psfrag{Frequency of viewings along the week for users of household 200}[bc][tc]{{\scriptsize household 200}}
\psfrag{Frequency of viewings along the week for users of household 203}[bc][tc]{{\scriptsize household 203}}
\psfrag{Frequency of viewings along the week for users of household 266}[bc][tc]{{\scriptsize household 266}}
\setlength{\unitlength}{0.1\textwidth}
\begin{picture}(5,4)
\put(0,2){\includegraphics[height=0.20\textwidth,width=0.26\textwidth]{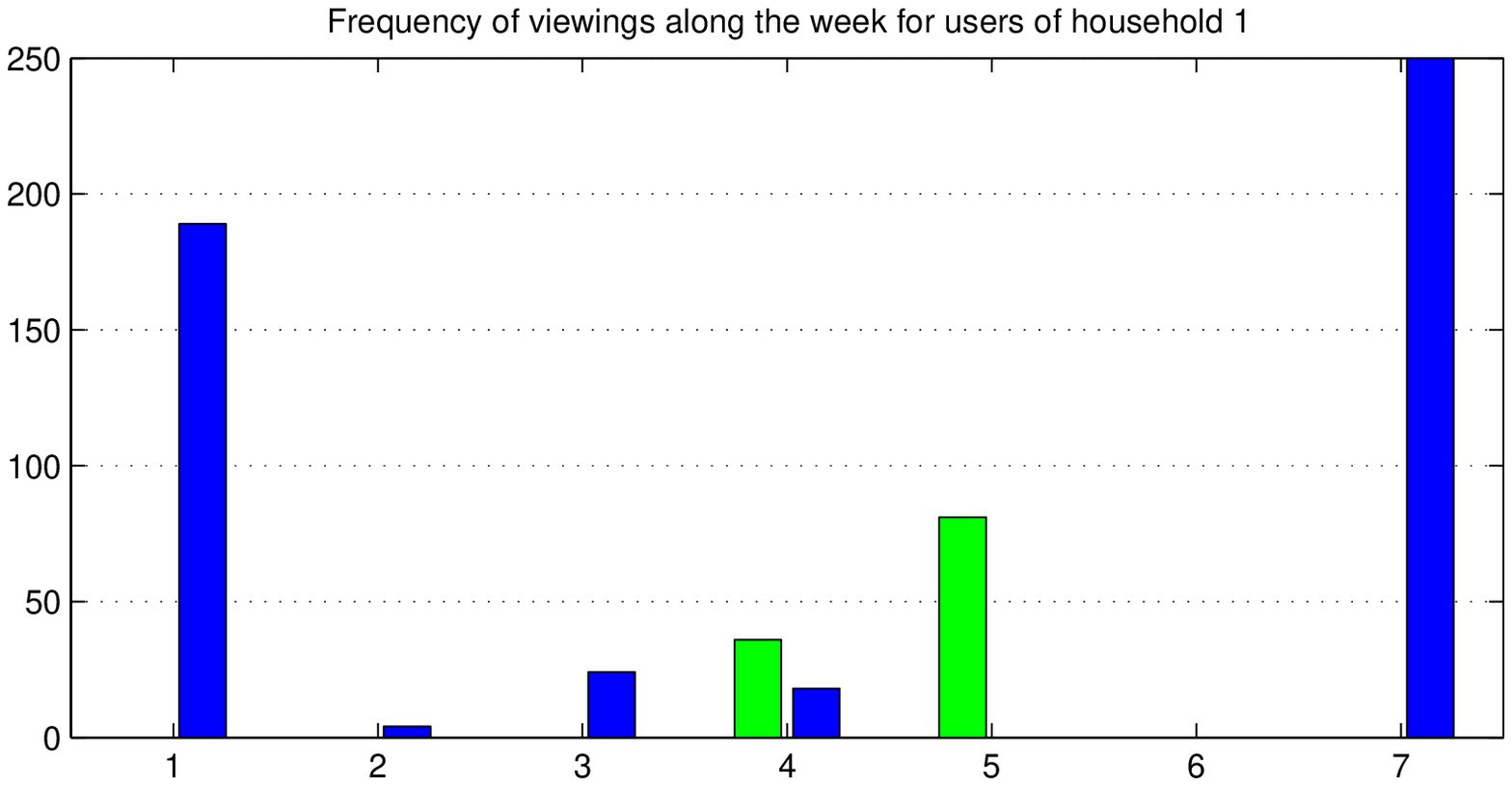}}
\put(2.5,2){ \includegraphics[height=0.20\textwidth,width=0.26\textwidth]{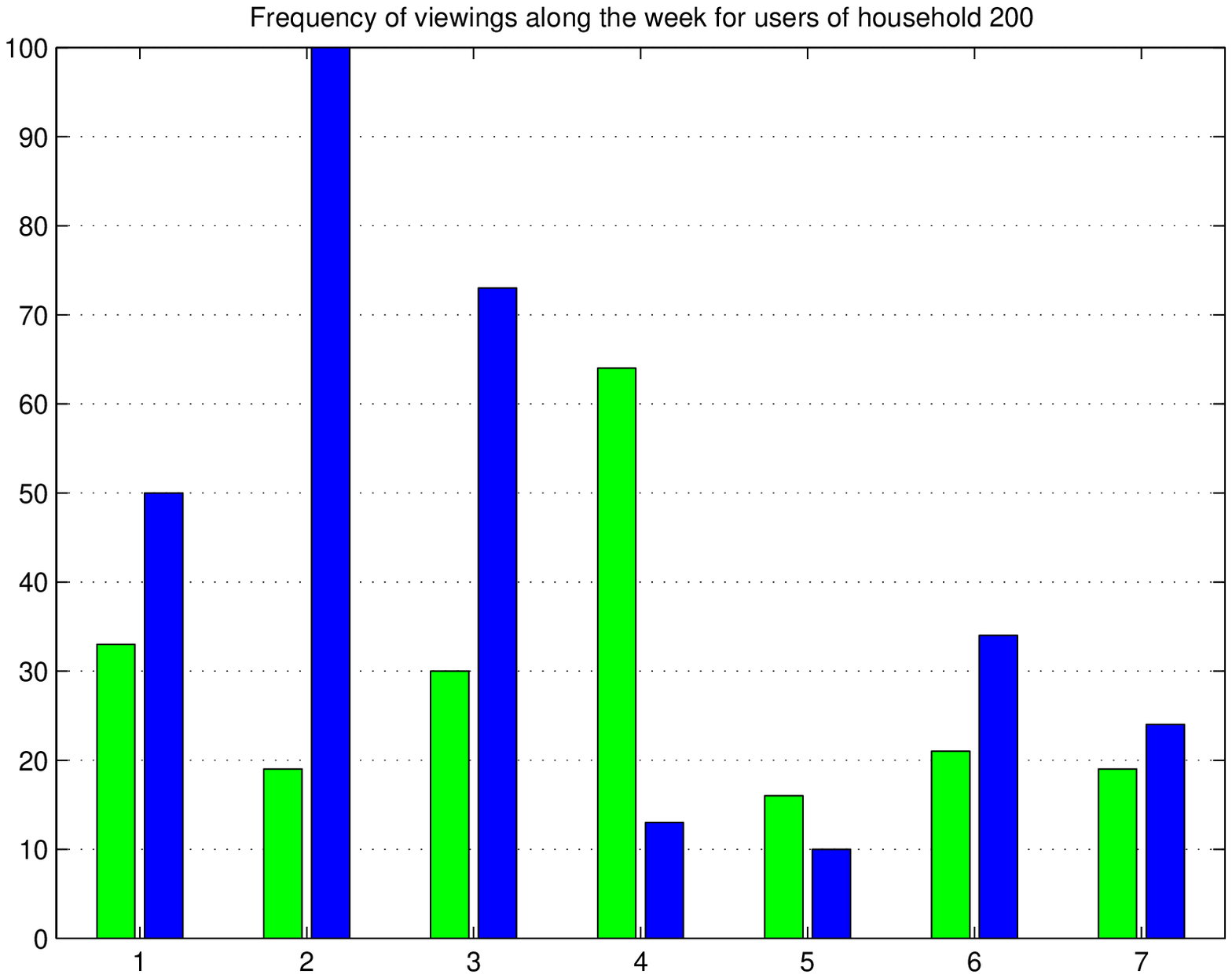} }
\put(0,0){ \includegraphics[height=0.20\textwidth,width=0.26\textwidth]{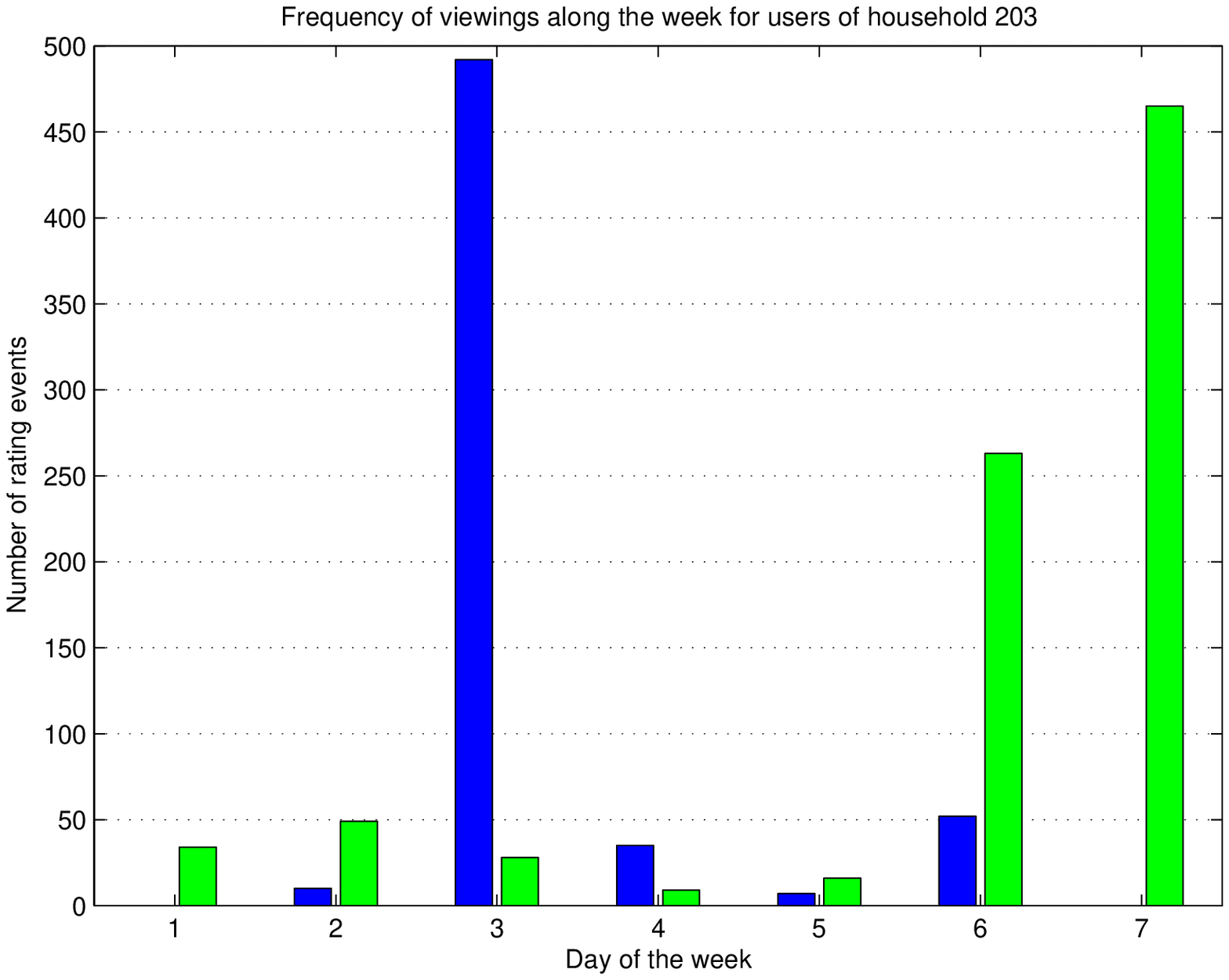} }
\put(2.5,0){  \includegraphics[height=0.20\textwidth,width=0.26\textwidth]{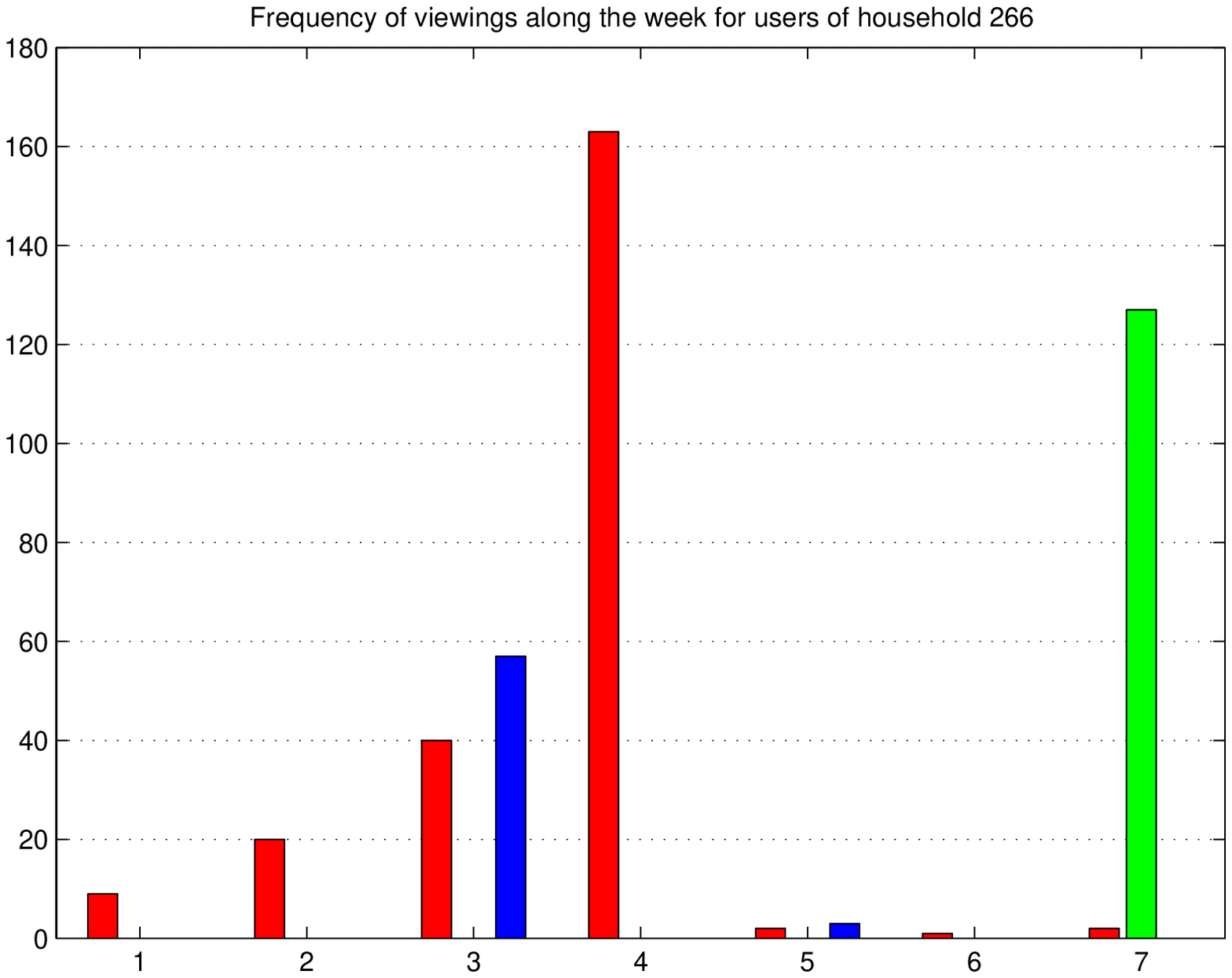} }
\end{picture}
\caption{Histograms of rating events across days of the week (day 1 is Sunday) for four
  households. The first three households have two members, while the
  fourth has three. For each day of the week, we plot $|H|$ histograms
  in different colors, each indicating the number of viewing events of
  a household member.}\label{fig:homes}
\end{figure}

As an illustration, Figure~\ref{fig:homes} shows the frequencies
with which users view movies on different days of
the week for four  households (labeled $1$, $200$, $203$, and $266$ in the
training set). We see that, in households $1$, $203$, and $266$, household
members tend to view and rate movies at very distinct days of the week.
For example, in household $1$,
one user watches movies mostly on Sunday and Saturday, while the
other watches movies in the middle of the week.
%{\bf [A: Why in the plot for household 266 there is no box?]}

This  phenomenon is  repeated in most of the households in the
training set. In order to quantify our observation, let $p_i(d)$
denote the empirical probability distribution  of rating events
associated with user $i\in [m]$ over different days $d\in
\mathcal{W}=\{\text{Sun},\text{Mon},\ldots, \text{Sat}\}$
(normalized so that $\sum_{d\in\mathcal{W}}p_i(d)=1$). We define the average total variation of a household $H$ as
$$\delta_H=\frac{1}{|H|(|H|-1)}\sum_{i,i'\in H} \|p_i-p_{i'}\|_{TV},$$
where we recall that $\|p-q\|_{TV}=\sum_{d\in\mathcal{W}}\frac{1}{2}|p(d)-q(d)|$.
By definition  $\delta_H\in [0,1]$, with $\delta_H=1$ corresponding to
a household in which no two users both rated a movie on the same day
of the week (possibly in different weeks).

Figure~\ref{tvpdf} shows the empirical probability distribution of
$\delta_H$ across different households $H$.
The distribution of $\delta_H$ is well concentrated around $1$, with
more than $70\%$ having  $\delta_H>0.8$. This is a quantitative
measure of the phenomenon suggested by Figure \ref{fig:homes}.

\subsection{Viewer prediction based on time-stamps}\label{sec:timeprediction}
\begin{figure}[t!]
\setlength{\unitlength}{0.1\textwidth}
\begin{picture}(5,3.5)
\put(0,0){\includegraphics[height=0.35\textwidth,width=0.50\textwidth]{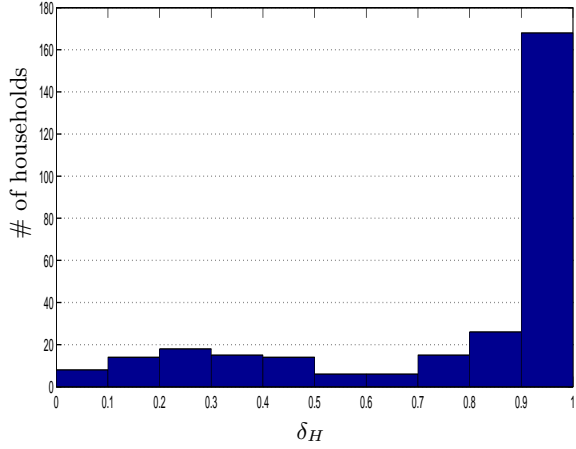}}
\put(0.3,1.5){\begin{sideways}\# of households\end{sideways}}
\put(2.45,0){$\delta_H$}
\end{picture}
\caption{Histogram of the average total variation distance $\delta_H$
  across the 290 households
in the training dataset. The majority of households have an average total variation close to 1, indicating that the distributions of rating events by different household members have almost disjoint supports.}
\label{tvpdf}
\end{figure}

 In
this section, we present three simple predictors of the household
member who watches a movie. Our third predictor exploits the fact that the day of the week can serve as a
very good indicator of which member is  watching a movie, as suggested by Figure \ref{tvpdf}.
%None of these three predictors takes into account the actual rating given by a user.
Our predictors maximize the likelihood a given member rated a movie; each predictor assumes a different model of how movie ratings take place. 

\sloppy
The simplest model assumes that each time a movie is watched in
household $H$, the user $i\in H$ is chosen at random with distribution
$q_{H}(i)$ independent of everything else.
This probability can be estimated  from the training set as follows
for household $H$
(we suppress the household subscript since this is fixed to $H$ throughout):
\begin{align*}
q(i) & = \frac{|\{(i',j,M_{i'j},t_{i'j})\in \Train:
  i'=i\}|}{|\{i',j,M_{i'j},t_{i'j})\in \Train: i'\in H\}|}\, .
\end{align*}
Given a time $t$ at which a movie is viewed, recall that $b(t)\in
\{1,\ldots,T\}$ denotes the time bin. As in the previous
section, we use $T=12$ here (one bin per month).
In the second model, the probability that the rating was given by user
$i$ depends only on the time bin $b(t)$ in which it occurred, and is
independent from everything else, conditional on $b(t)$: 
\begin{align*}
q(i\mid b(t))  & = \frac{|\{(i',j,M_{i'j},t_{i'j})\!\in\! \Train: i'\!=\!i
  \land b(t_{i'j}) \!=\! b(t)\}|}
{|\{i',j,M_{i'j},t_{i'j})\!\in\! \Train: i'\!\in\! H \land b(t_{i'j})\! =\!
  b(t)\}|}.
\end{align*}
Finally, let $d(t)\in \mathcal{W} = \{\text{Sun},\text{Mon},\dots \text{Sat}\}$ be the day of the
week at which the viewing occurs. Our
third model assumes that the user who rated the movie is independent from
everything else, conditional on the day of the week:
\begin{align*}
q(i\mid d(t))  & = \frac{|\{(i',j,M_{i'j},t_{i'j})\!\in \!\Train: i'\!=\!i
  \land d(t_{i'j}) \!=\! d(t)\}|}{|\{i',j,M_{i'j},t_{i'j})\!\in\! \Train:
  i'\!\in\! H \land d(t_{i'j}) \!=\!
d(t)\}|}\,.
\end{align*}
Given a tuple $(H,j,M_{Hj},t_{Hj})\in \Test$, we can consider the following three simple classification algorithms:
\begin{align*}\underset{i\in H}{\mathop{\arg\!\max}\,} &q(i), & \underset{i\in H}{\mathop{\arg\!\max}\,} q(i\mid b(t_{Hj}))&, & \underset{i\in H}{\mathop{\arg\!\max}\,} q(i\mid d(t_{Hj}))&.\end{align*}
Note that the second and third algorithms make use of the time at
which a viewing event takes place. None of the three uses the actual rating
$M_{Hj}$ given by the user. We present an algorithm that does use the rating in the next section.
\fussy

\subsection{Generative model}\label{sec:generative}

\begin{figure}[t]
%\centering
\setlength{\unitlength}{0.1\textwidth}
\begin{picture}(5,2)
\put(-0.4,0.1){
\includegraphics[height=0.2\textwidth,width=0.28\textwidth]{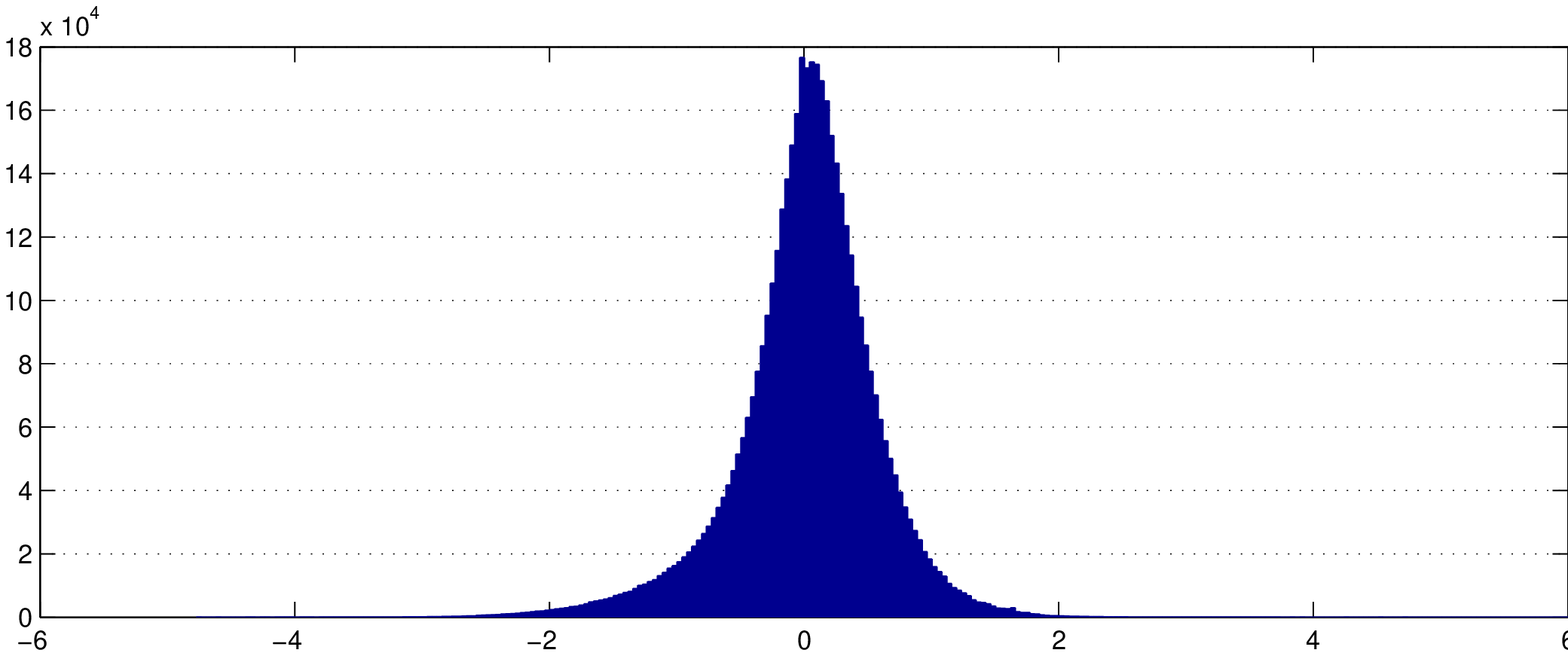}
}
\put(2.3,0.1){
\includegraphics[height=0.2\textwidth,width=0.28\textwidth]{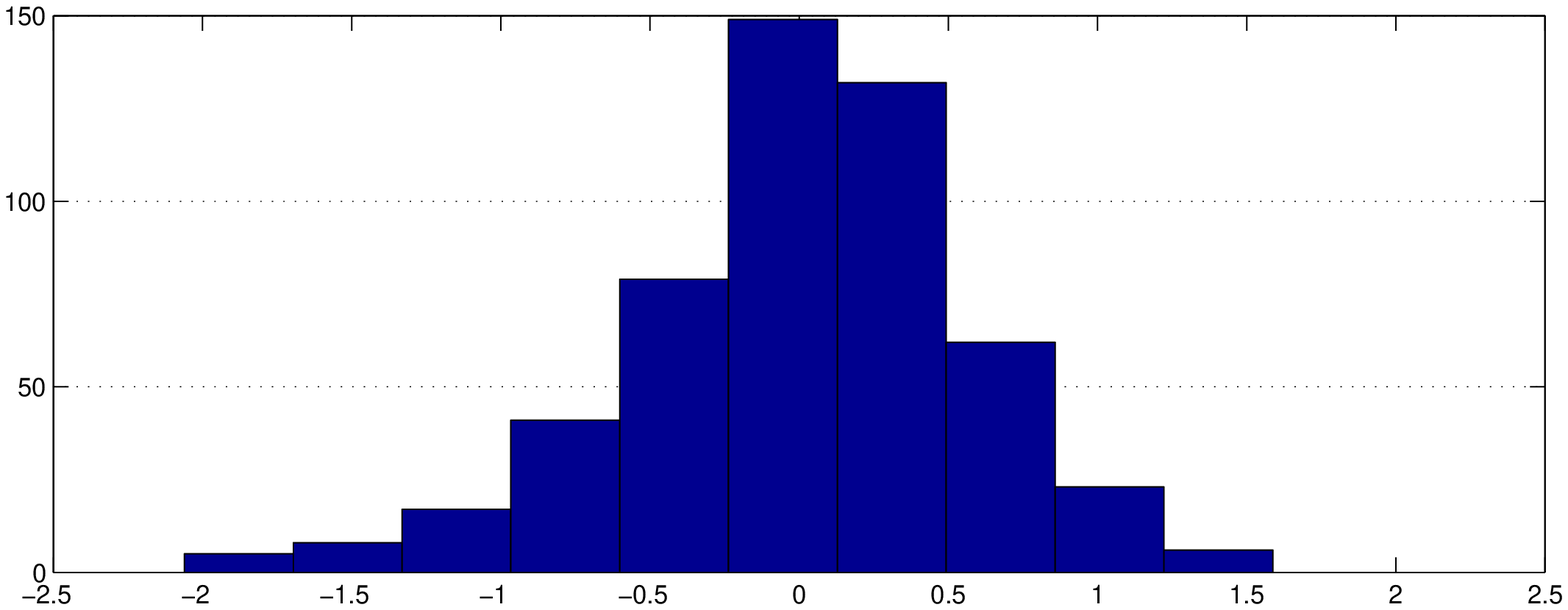}
}
\put(0.7,0){\small (a) All ratings}
\put(2.7,0){\small (b) Ratings by single user}
\end{picture}
\caption{PDF of the residual error across (a) all ratings in the training dataset and (b) all ratings given by a single user. The distributions are well approximated by normals.}\label{residual}
\end{figure}

In order to account for ratings given by the users in our prediction,
we introduce a generative model for how users rate movies. Our model
assumes that the rating given by a user is normally distributed around
the prediction made by the low rank approximation algorithm of
Section~\ref{sec:LowRank}.
In particular, recall that the predicted rating of a user $i\in [m]$ viewing movie $j\in [n]$ at time $t$ is given by
\begin{align}\label{pred}\hM_{ij}(b(t)) = z_{i}(b(t))+\langle u_i(b(t)),v_j(b(t)) \rangle \end{align}
where $u_i$, $v_j\in \reals^r$ are the vectors associated with $i$ and
$j$, respectively, and $z_{i}$ is the centering component.
This prediction depends on the time-stamp $t$ only through the bin
$b(t)$.
Figure~\ref{residual}(a) shows the distribution of the residual error
$$M_{ij} - \hM_{i,j}(b(t_{ij})) $$
across all user/movie pairs $(i,j)$ in the training set. The
distribution seems to be well approximated by a normal distribution,
Figure~\ref{residual}(b) shows the distribution of residuals for a
single user  (user with ID 56094~in the training set). This still roughly agrees with a Gaussian distribution,
although not as closely as for the overall distribution.

This motivates modeling the rating given by a user $i$ for a movie $j$ at time $t$ by a normal distribution $\normal(\hM_{ij}(b(t)),\sigma)$, where $\hM_{ij}(b(t))$ is given by \eqref{pred}  and $\sigma^2$ is the variance of the residual error, as estimated from the training set.
More specifically, given that a user from household $H$ views a movie
$j$ at time $t_{Hj}$, we model the joint probability that $(a)$
user $i\in H$ is the rater and  $(b)$ $i$ gives a rating $M$ as follows:
\begin{align}
\prob(i,M) =
\frac{1}{S}\,e^{-\frac{\left(M-\hM_{ij}(b(t_{Hj}))\right)^2}{2\sigma^2}}\,
q(i)\, .\label{joint}
\end{align}
where $S\equiv \sqrt{2\pi \sigma^2}$.
Alternative models are obtained if we
condition on the bin or the day of the rating, as discussed in the
previous section:
\begin{align}
\prob(i,M\mid b(t_{Hj}))& =
\frac{1}{S}\,e^{-\frac{\left(M-\hat{M}_{ij}(b(t_{Hj}))\right)^2}{2\sigma^2}}\, q(i\mid b(t_{Hj})),\label{jointmidbin}\\
\prob(i,M\mid d(t_{Hj})) &=
\frac{1}{S}\,e^{-\frac{\left(M-\hat{M}_{ij}(b(t_{Hj}))\right)^2}{2\sigma^2}}\, q(i\mid d(t_{Hj})).\label{jointmidday}
\end{align}

Given a tuple $(H,j,M_{Hj},t_{Hj})\in\Test$, the posterior probability that $i\in H$ is the movie viewer under the above three generative models can be written as:
$$ \prob(i\mid M_{Hj},\;\cdot\;) = {\prob(i, M_{Hj}\mid
  \;\cdot\;)}/{\sum_{i'\in H}\prob (i', M_{Hj}\mid \;\cdot\;)}.$$
As a result, the following rule can be used as a classifier of tuples $(H,j,M_{Hj},t_{Hj})\in\Test$:
$$\underset{i\in H}{ \mathop{\arg\!\max}}\, \prob(i, M_{Hj}\mid \;\cdot\;) $$
where $\prob(i, M_{Hj}\mid\;\cdot\; )$ is given for each of the three
generative models by \eqref{joint}, \eqref{jointmidbin} an
\eqref{jointmidday}, respectively.
%
%******************************************
%
\subsection{Empirical results}
%\begin{table}

We evaluated the classification algorithms of
Sections~\ref{sec:timeprediction} and~\ref{sec:generative} by
cross validation on the training and test sets, as described in Section
\ref{sec:LowRankParam}. For classifiers based on the
generative models of  Section~\ref{sec:generative}, the  low-rank
model was selected to be the same as in Section
\ref{sec:LowRankParam}
(in particular we used $T=12$, $r=10$,
$\lambda=1$, $\xi_u=10$, $\xi_v=\xi_z=40$).

%{\bf [A:Please fill in the above]}

\begin{table}
\begin{small}
\begin{tabular}{r|ccc}
 & $\sigma=\infty$ & $\sigma=\sigma_{\text{all}}$ & $\sigma=\sigma_i$  \\
\hline
$q(i)$ & $0.3916\!\pm\!0.0081$ & $0.3264\!\pm\!0.0102$ & $0.3066\!\pm\!0.0112$\\
$\!\!q(i\!\mid\! b(t_{Hj}))$ & $0.3626\!\pm\!0.0080$
 &  $0.2956\!\pm\!0.0065$ &  $0.2777\!\pm\!0.0084$ \\
$\!\!q(i\!\mid\! d(t_{Hj}))$& $0.1129\!\pm\!0.0066$ & $0.1008\!\pm\!0.0066$ & $0.0966\!\pm\!0.0072$\\
\hline
\end{tabular}
\end{small}
\caption{Misclassification rates $\P$ for algorithms of Sections~\ref{sec:timeprediction} and~\ref{sec:generative}, with standard deviations derived over five iterations of cross validation.}\label{table:misclass}
\end{table}

The results are summarized in Table~\ref{table:misclass} in terms of the misclassification rate. The first column of the table
($\sigma=\infty$) corresponds to the classifiers of Section~\ref{sec:timeprediction} (not using the ratings). The second
and third columns correspond to the classifiers
outlined in Section~\ref{sec:generative}. In the second column, the
variance $\sigma$ used in the normal distribution is
estimated by the empirical variance of the residual errors over all
ratings in the training set. In the third column, we used a
user-dependent variance $\sigma_i$ for each $i\in [m]$.
This is estimated by the variance of the residual errors of ratings
given by $i$. Finally, each row corresponds to a
 different  assumption on the posterior probability $q$, with the second and third rows corresponding to the use of bin and weekday information, respectively (\emph{c.f.}~Eq.~\eqref{jointmidbin} and ~\eqref{jointmidday}).

We observe that, in all cases, using the bin information helps
compared to using the unconditional probability $q(i)$, but only
marginally so. The largest improvement comes from conditioning on the
day of the week. This  decreases the misclassification rate by a factor between 3 and 4 compared to using the unconditional probability $q(i)$. Incorporating the generative model also decreases the misclassification rate:
 classification using the generative model conditioned on the day of
 the week, along with individual variances $\sigma_i$, outperforms all
 other methods, with $\P\approx 0.0966$.

As mentioned above, these are misclassification rates estimated
through five-fold cross-validation. We report these in detail because
they provide a metric that is statistically more robust.
When using the original split in train and test sets provided in the challenge,
we achieve (for the third column, $\sigma=\sigma_i$)
respectively $\P\approx 0.3028$ (model $q(i)$), $0.2765$ (model
$q(i|b(t_{Hj}))$), $0.0950$ (model $q(i|d(t_{Hj}))$). For this same split,
and for the model $q(i|d(t_{Hj}))$, the values for $\P_2$, $\P_3$
and $\P_4$ are $0.0940$, $0.1051$ and $0.1315$ respectively.

Finally, these results remain excellent if evaluated in terms of ROC
curves, and Area Under the Curve ($\AUC$). We compute $\AUC$ as follows. Consider a household $H$,  a
user $i$, and the corresponding probabilities $p_j = \prob(i\mid
M_{Hj},\;\cdot\;)$. Let $a$ be the number of unordered pairs $(j, j')$ such that $p_j>p_{j'}$ and $j'$ was indeed rated by $i$, while $j$ was not. Let $b$ be the product
between the number of entries in the test set that were rated by user $i$ and the number of entries that were not. Define $\AUC_{i,H} = 1 - a/b$. $\AUC_{i,H}$ is the area
under the ROC curve for user $i$ versus any other user in household $H$.
We estimate $\AUC$ by averaging the above quantity over $i$ and  $H$
in the test set for which $b \neq 0$.
Using the original split in test and train set provided with the challenge dataset, we
obtain
(again for the third column, $\sigma=\sigma_i$)
respectively $\AUC\approx 0.6170$ (model $q(i)$), $0.6619$ (model
$q(i|b(t_{Hj}))$), $0.8947$ (model $q(i|d(t_{Hj}))$).
%
%=========================================================================
%
\section{A unified framework}
\label{sec:Unified}

%\begin{figure}[t]
%\includegraphics[width=0.98\columnwidth]{LR-PredictionError.eps}
%\caption{Average prediction error vs. number of iterations for different sets of parameters}
%\label{fig:LogRec-PredErr}
%\end{figure}

%\begin{figure}[t]
%\includegraphics[width=0.98\columnwidth]{LR-ROC.eps}
%\caption{Average prediction error vs. number of iterations for different sets of parameters}
%\label{fig:LogRec-PredErr}
%\end{figure}

While the generative models studied in the previous section yield
excellent results, it is possible to improve upon them by including further
contextual information. As an example, the rating time-stamp also
provides us information on the time of the day at which the rating was
entered. In many households, the separation of temporal patterns
discussed in Section \ref{sec:temporal} becomes more acute when
including the time of the day.
This raises the need of developing a
systematic scalable way of incorporating such information.

Our approach is to formulate the problem as a supervised multinomial classification
problem. The challenge of constructing a classifier can then be decoupled
in two separate two sub-tasks: $(i)$ Constructing a generic
multinomial classifier (or choosing one from the vast literature on
this topic); $(ii)$ constructing a suitable set of features.

In order to illustrate this approach, we describe it for
a deliberately simple classifier: $\ell_1$-regularized logistic
regression.  Furthermore, we reduce the classification problem to a binary one.
Fix a household $H$, and a user $i\in  H$ (omitting hereafter
reference to $i$ and $H$ whenever possible). Each rating event within
household $H$ is then characterized by the pair $(y,\cO)$. Here
$y$ is a binary variable, equal to $1$ if and only if the rating was
provided by $i$, and $\cO$   denotes collectively the other available
information about the event.
We then assume a logit model
\begin{eqnarray}
\prob(y=1|\cO) = \frac{e^{\<\theta,x(\cO)\>}}{1+e^{\<\theta,x(\cO)\>}}\, ,\label{eq:LogitProb}
\end{eqnarray}
whereby $x(\cO)\in\reals^p$ is a feature vector constructed from the
available information, and $\theta = \theta_{i,H}\in \reals^p$ is a
vector of parameters to be fitted from the data.
Assuming the parameters are known, a rating will be attributed to user
$i$ if this maximizes the probability  (\ref{eq:LogitProb}) among all
the users in the same household.

In order to learn the parameters $\theta = \theta_{i,H}$, we consider
the training rating events within household $H$, and index them by $s\in
\{1,\dots,N_H\}$. Denoting the $s$-th such event by $(y_s,\cO_s)$, we
consider the regularized likelihood
\begin{align*}
\cL(\theta) \equiv
-\sum_{s=1}^{N_H}\big\{y_s\<\theta,x(\cO_s)\>-\log\big(1+e^{\<\theta,x(\cO_s)\>}\big)+\lambda_1\|\theta\|_1\, .
\end{align*}
Once again we emphasize that regularized logistic regression is not
necessarily the best classification method, and our approach accommodates
alternative algorithms.

We implemented this procedure using \rm{l1logreg}, a
software that minimizes  $\cL(\theta) $ based on an interior
point method described in \cite{ZKOH07}. All the data was standardized
before being introduced into the solver. The algorithm was tested for
different feature vectors constructed by including at most the following:

\vspace{0.1cm}

\noindent $(a)$ The day of the week of the rating (i.e. $d(t_{ij})$)
implemented as a length-$7$ binary indicator vector.

\vspace{0.1cm}

\noindent $(b)$ The hour of the day of the vector,
implemented as a length-$24$ binary indicator vector.

\vspace{0.1cm}

\noindent $(c)$ The movie feature vector\footnote{The misclassification rate $\P \approx 0.37$ obtained using the low-rank model in Section \ref{sec:LowRankParam} can be lowered to $0.30$ by binning the time-stamps into 7 different bins, one per day of the week. This suggests adopting a 7-bin model of vectors on a per week-day (rather than per month) basis. However, adopting a 7-bin model did not improve the performance of the other classification algorithms introduced in  the paper, which rely on and outperform the low-rank model. This is also the case when, in the unified framework described in this section, we include in $x(\cO)$ the vector $v_j(d(t_{Hj}))$ instead of $v_j(b(t_{Hj}))$.} $v_j(b(t_{ij}))\in \reals^r$, learned from
the low-rank model of Section \ref{subsec:LowRankTD}.

\vspace{0.1cm}

\noindent $(d)$ The time bin $b(t_{Hj})$ implemented as a length-$12$ binary indicator vector.

\vspace{0.1cm}

\noindent $(e)$ The actual rating $M_{ij} \in \{0,...,100\}$ scaled and shifted
so that $0$ corresponds to $1$ and $100$ to $5$.

\vspace{0.1cm}

Table \ref{table:misclassRLR}
shows how we reach our best values for $\P$ as we include more and more
features in the feature vector. Although when doing cross validation including
more feature seems to help, for the challenge test set, not including the
rating produces best results. We note however that the way we are using
the regularized logistic regression can be easily improved by assigning
different regularization weights to different components of the feature vector (right now we are using the same weight, $\lambda_1$). This might explain why
including certain features is not improving the results.

With this choice of $x(\cO)$, and $\lambda_1 = 0.01$,
we achieved misclassification rate $\P=0.0419 \pm 0.0026$ and area under the curve
$\AUC = 0.9689 \pm 0.0027$, as estimated through the subsampling procedure described
above. On the challenge test set, and not including the ratings in $x(\cO)$, the same performance metrics
evaluated to $\P=0.0406$ and $\AUC = 0.9611$. 

For the challenge test 
set the values of $P_2$, $P_3$ and $P_4$ are $0.0413$, $0.0268$ and $0.0463$ respectively. We note that the misclassification rate is smaller for households
with 3 users. This is contrary to the natural intuition that the more people belong to a household the harder it should be to distinguish between them.

\begin{table}
\begin{small}
\begin{tabular}{r|ccc}
 & 5 fold cross validation & Challenge test set  \\
\hline
$(a)$ & $0.1137 \pm 0.0077$ & $0.1142$\\
$(a),(b)$ & $0.0483 \pm 0.0039$ & $0.0570$\\
$(a),(b),(c)$ & $0.0468 \pm 0.0032$ & $0.0463$\\
$(a),...,(d)$ & $0.0423 \pm 0.0020$ & $\mathbf{0.0406}$\\
$(a),...,(e)$ & $0.0419 \pm 0.0026$ & $0.0412$\\
\hline
\end{tabular}
\end{small}
\caption{Misclassification rates $\P$ using the regularized logistic regression for $\lambda_1 = 0.01$ and sequentially
including more features into the feature vector. The performance of our best predictor on the challenge test set is noted in bold.}\label{table:misclassRLR}
\end{table}

\bibliographystyle{abbrv}

\begin{small}
\bibliography{CAMRaCameraReady}
\end{small}

% sigproc.bib is the name of the Bibliography in this case
% latex bibtex latex latex/
% ACM needs 'a single self-contained file'!
%
\newpage

\end{document}